\documentclass[aps,pra,twocolumn,amsmath,amssymb,superscriptaddress]{revtex4-2}

\usepackage{comment}

\usepackage{amsmath,amsfonts,amssymb}
\usepackage{physics}
\usepackage{graphicx}
\usepackage{setspace}
\usepackage{tocloft}
\usepackage[dvipsnames]{xcolor}

\cftpagenumbersoff{figure}
\cftpagenumbersoff{table} 

\usepackage{CJK}
\usepackage{mathtools}
\usepackage[normalem]{ulem}
\usepackage{dcolumn}
\usepackage[mathlines]{lineno}
\usepackage{hyperref}
\usepackage{bm}
\usepackage{amssymb}
\usepackage{amsmath}
\usepackage{braket}
\usepackage{color}
\usepackage{xcolor}
\usepackage{soul} 
\usepackage{graphics}
\usepackage{bbm}

\begin{document} 
\title{Topological robustness of optical skyrmions through a real-world free-space link}

\author{Cade Peters}
\affiliation{School of Physics, University of the Witwatersrand, Private Bag 3, Johannesburg 2050, South Africa}

\author{Vagharshak Hakobyan}
\affiliation{University of Bordeaux, CNRS, Laboratoire Ondes et Mati\`{e}re d'Aquitaine, Talence, France}

\author{Alice Drozdov}
\affiliation{School of Electrical and Information Engineering, University of the Witwatersrand, Private Bag 3, Johannesburg 2050, South Africa}

\author{Etienne Brasselet}
\affiliation{University of Bordeaux, CNRS, Laboratoire Ondes et Mati\`{e}re d'Aquitaine, Talence, France}

\author{Mitchell Cox}
\affiliation{School of Electrical and Information Engineering, University of the Witwatersrand, Private Bag 3, Johannesburg 2050, South Africa}

\author{Andrew Forbes}
\email[email:]{ andrew.forbes@wits.ac.za}
\affiliation{School of Physics, University of the Witwatersrand, Private Bag 3, Johannesburg 2050, South Africa}
\email[Corresponding author: ]{andrew.forbes@wits.ac.za}

\date{\today}

\begin{abstract}
\noindent  \textbf{Structured light offers a promising solution for the increasing data demands of modern optical networks, opening up new degrees of freedom that can be leveraged for greater channel capacity and more bits per photon. However, its implementation is hindered by real-world distortions, for example, atmospheric turbulence in free-space, with severe and rapidly evolving phase perturbations that alter the amplitude, phase and vectorial polarization structure of the beam. Here, we demonstrate that optical topologies in the form of skyrmions are highly resilient to the effects of real-world atmospheric turbulence. We create and transmit these particle-like topologies of light through a 270~m free-space optical link, revealing their robustness across a wide variety of conditions and turbulence strengths. While we observe severe distortion in the states' underlying degrees of freedom, we show that the topological numbers are preserved in all cases. We account for fast changes to the medium, where the channel produces statistically averaged outcomes, by probing the state's decoherence, showing that while the degree of polarisation consequently decays, the topology remains intact. Using topology, we show information can be transmitted through the channel with almost perfect fidelity ($>98\%$) in most cases, only decreasing to $86\%$ in the most severe conditions tested. Our work is the first to demonstrate the potential for optical topologies as reliable and robust information carriers in a real-world environment and points to the potential for other complex channels too, offering attractive features for classical and quantum communication alike.}
\end{abstract}

\maketitle

\noindent The use of light to transmit information has had a profound impact on how we communicate. The ever increasing need for higher transmission speeds and greater bandwidths has driven a wide variety of research efforts, with novel encoding schemes that involve multiplexing in time \cite{richter2011transmission}, wavelength \cite{sano2011ultra}, polarisation \cite{zhou201164} and combinations of these \cite{gnauck2011spectrally,richardson2010filling}, allowing for significant advancements in modern communication networks. Free space optical communications (FSOC) has especially attracted interest, with the ability to establish high speed, secure and licence-free communications links that are energy efficient and easy to deploy \cite{lavery2018tackling,trichili2020roadmap}. Structured light \cite{forbes2021structured} offers the potential to further increase the capacity of these channels through the use of space-division \cite{liu20111,li2014space,richardson2013space} and mode-division multiplexing \cite{berdague1982mode,gibson2004free,djordjevic2011deep} enabled by the tailoring of multiple degrees of freedom (DoFs) in both classical and quantum regimes \cite{rubinsztein2016roadmap,nape2023quantum,shen2021rays}.

The primary factors limiting the use of structured light in these contexts are the deleterious effects of the complex phase profiles and optical scatterers present in the real-world media through which light propagates. This includes biological tissue \cite{yoon2020deep,He2021polarisation,ntziachristos2010going}, optical fibre \cite{cao2023controlling}, and most importantly for FSOC: atmospheric turbulence \cite{peters2025structured}. Non-uniform refractive index profiles caused by variations in temperature and pressure result in unwanted coupling within these high dimensional DoFs in classical links \cite{anguita2008turbulence,krenn2014communication,rodenburg2012influence,krenn2016twisted,zhang2020mode,ren2016experimental,paterson2005atmospheric,cox2019resilience} and a loss of entanglement for quantum channels \cite{jha2010effects,hamadou2013orbital,gopaul2007effect,pors2011transport,leonhard2015universal,sorelli2019entanglement} resulting in decreased performance and the loss of information.

Strategies to mitigate these effects have taken on many forms, including the use of adaptive optics \cite{tyson2002bit,zhao2011aberration,ren2014adaptive}, machine learning \cite{liu2019deep}, modal diversity \cite{cox2018modal} and phase conjugation with non-linear optics \cite{singh2024light,zhou2025automatic} to name a few. An alternative and exciting approach is to leverage the physics of atmospheric turbulence to find or create invariant forms of structured light, that can pass through medium as if it were transparent. Such invariants can take many forms with recent studies including eigenmodes \cite{klug2023robust,peters2025tailoring}, optical knots \cite{pires2025stability} and the non-separability of the vectorial beams \cite{nape2022revealing,peters2023invariance} each with their own benefits and limitations. Optical skyrmions \cite{shen2024optical} are easy to generate, scale and measure and bring to bear the many advantages of topology, where the mapping from real-space to a given parameter space is preserved, so long as the channel can be described as a smooth deformation of the map. It is not yet know which channels meet this criterion and finding the channels that do is an active area of research. It has recently been demonstrated in controlled laboratory studies that classical skyrmion are robustness to known optical transformations  \cite{wang2024topological}, which has been extended to scattering media \cite{peters2025seeing} and theoretically simulated atmospheric turbulence \cite{wang2025robustness}. Likewise, it has also been demonstrated that non-local quantum skyrmions are immune to the effects of entanglement decay \cite{ornelas2024non} and isotropic noise \cite{ornelas2025topological}. However, there has yet been no demonstration of the robustness of any optical topology through realistic environments.

Here, we present results testing the stability of topological light through a real-world atmospheric channel. We prepare several optical skyrmions and transport them through a 270~m free space optical link located on the Braamfontein campus of the University of the Witwatersrand, in the centre of Johannesburg, South Africa. We measure the topology with a simple, single shot, spatially resolved Stoke polarimetry setup, allowing for the simultaneous measurement of all four Stokes parameters. We demonstrate that the topological wrapping number remains robust through the channel over a wide variety of atmospheric conditions, ranging from calm, cool morning air to the highly erratic and intense distortions of the midday heat. We first present results of measurements made within the atmospheric coherence time $\tau_0$, the time scale at which the medium appears transparent to the propagating beam and the typical upper limit for most classical communication scheme. We show that in this regime, the topological wrapping number is highly robust, being easily recovered even when the amplitude and phase distortions induced by the channel leave the underlying beam unrecognizable and the vectorial polarization structure highly scrambled. Additionally, we also demonstrate the robustness of this topology to the decoherence and depolarization effects of turbulence, by averaging measurements over time scales greater than $\tau_0$, allowing the experienced distortion to evolve during the measurement. Such timescales are relevant for when long measurement times are needed, for example with slow detectors or when measuring single or biphoton states which require the build up statistics over time. Again, we show the preservation of the topological wrapping number despite an almost 40\% drop in the beams degree of polarization (DoP). All measurements are performed with no probing of the medium and no pre- or post-correction is made to compensate for the channel distortion. These results provide compelling evidence for the benefits of topological light and its potential to greatly improve the capacity of FSOC, both classical and quantum. They also motivate the development of new technologies for the effective and efficient generation and detection of topological light and open up new avenues for robust, correction free optical communication and transmission through complex channels 

\section*{Results}
 
\begin{figure}[h]
    \centering
    \includegraphics[width=\linewidth]{ 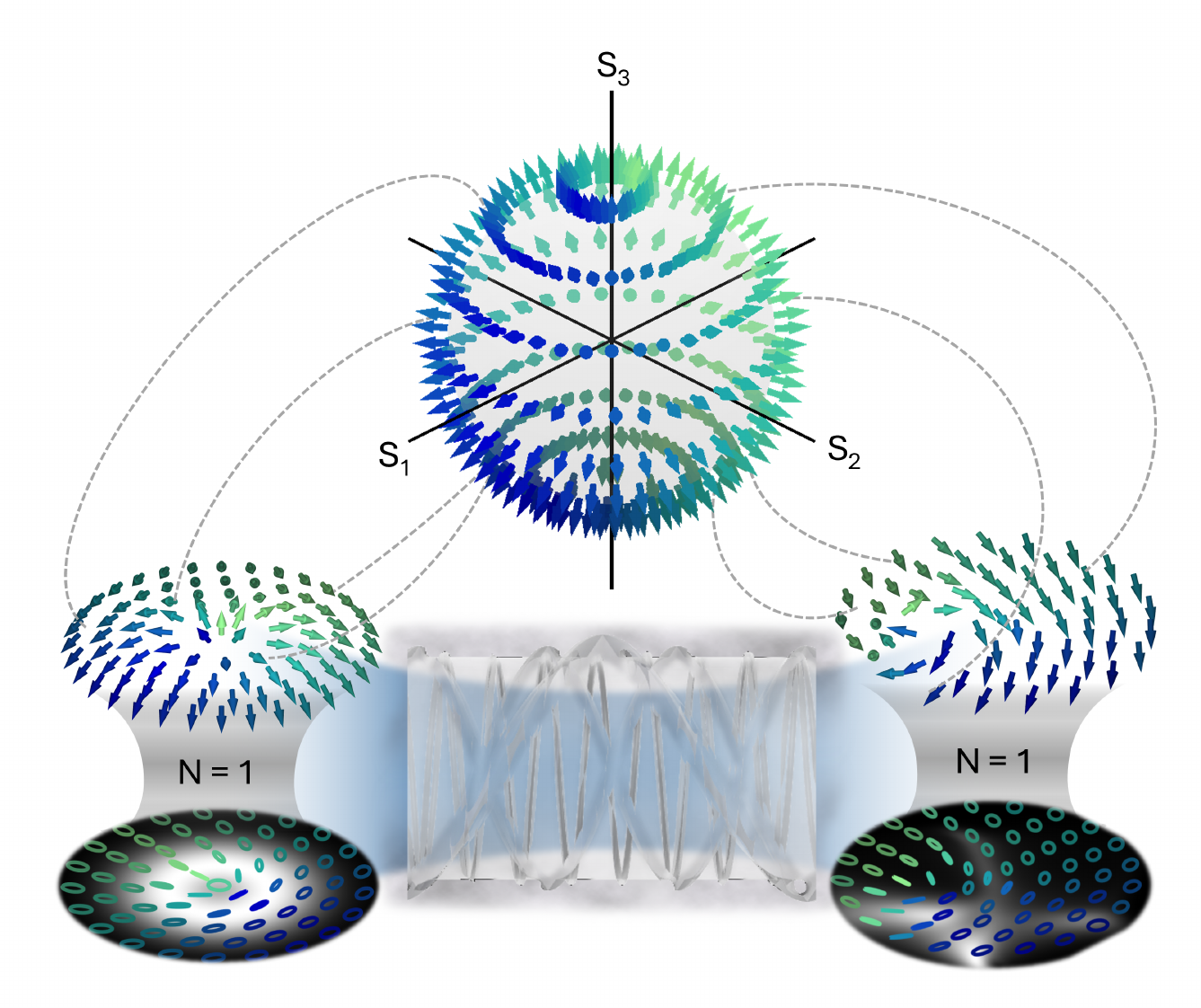}
    \caption{\textbf{Skyrmionic topology through turbulence}. Illustration showing how optical skyrmions are formed via a mapping from the transverse plane to the Poincar\'{e} sphere, characterised by the skyrmion number $N$. When these states pass through a turbulent channel, the beam's structure is heavily distorted in phase, amplitude and polarisation, but the mapping to the Poincar\'{e} sphere is unchanged.} 
    \label{fig:Concept}
\end{figure}
 
\noindent \textbf{Creation and detection of optical skyrmions} We construct optical skyrmions of the form
\begin{equation}
    \mathbf{U}(\mathbf{r}) = LG^0_{l_1}(\mathbf{r})\mathbf{e}_R + LG^0_{l_2}(\mathbf{r})\mathbf{e}_L \,,
    \label{eq:RL vector beam}
\end{equation}
where $LG_{l}^p(\mathbf{r})$ represents a Laguerre-Gaussian (LG) complex field, $\mathbf{r}$ is the transverse spatial coordinate and $\mathbf{e}_{R(L)}$ represent the right (left) circular polarisation unit vectors respectively. We set the radial index of the LG beam to $p=0$ for convenience and modulate the  azimuthal index $l$ to tune the topological wrapping number. This index imbues the beam with an orbital angular momentum (OAM) of $l\hbar$ per photon. States of this form are commonly referred to as vector beams \cite{zhan2009cylindrical,shen2022nonseparable} and exhibit complex, spatially varying polarisation structures across the transverse plane.
In the cases where $|l_1|\neq|l_2|$, the spatially varying polarisation structure, shown on the left hand side of Figure \ref{fig:Concept}, forms a skyrmionic topology. This mapping is characterised by a sphere to sphere mapping from the transverse plane $\mathcal{R}^2$ through a stereographic projection to the Poincaré sphere $\mathcal{S}^2$, shown by the mapping from the vector textured field in the transverse plane to the Poincar\'{e} sphere. Each skyrmionic topology is uniquely characterised by the 
topological wrapping number $N$, also called the skyrmion number. This integer value counts how many times one wraps the Poincar\'{e} sphere as one moves across the entire transverse plane \cite{shen2024optical}. The wrapping number for a skyrmion formed using the 3D Stokes vector field can be calculated as follows,
\begin{equation}
 N = \frac{1}{4\pi} \int_{\mathcal{R}^2}  \bm{S}\cdot\left( \frac{\partial  \bm{S}}{\partial x}\times  \frac{\partial  \bm{S}}{\partial y}\right)  \text{d}x\text{d}y \,,
    \label{eq:skyrmion wrapping}
\end{equation}
where $\bm{S}$ is the locally normalised Stokes vector  such that $\Sigma_{j=1}^3 S^2_j = 1$ to ensure the final mapping is always onto a unit sphere. For the above states, Equation \ref{eq:skyrmion wrapping}  simplifies to $N = n|l_1-l_2|$,
where $n =  \text{sign}(l_1-l_2)$. As illustrated in Figure \ref{fig:Concept}, when these states pass through atmospheric turbulence, the beam's amplitude profile and polarisation structure are severely distorted, but the mapping from the transverse plane to the Poincar\'{e} sphere, and consequently the topological wrapping number $N$, is maintained.

\begin{figure*}[htpb!]
    \centering
      \includegraphics[width=\linewidth]{ 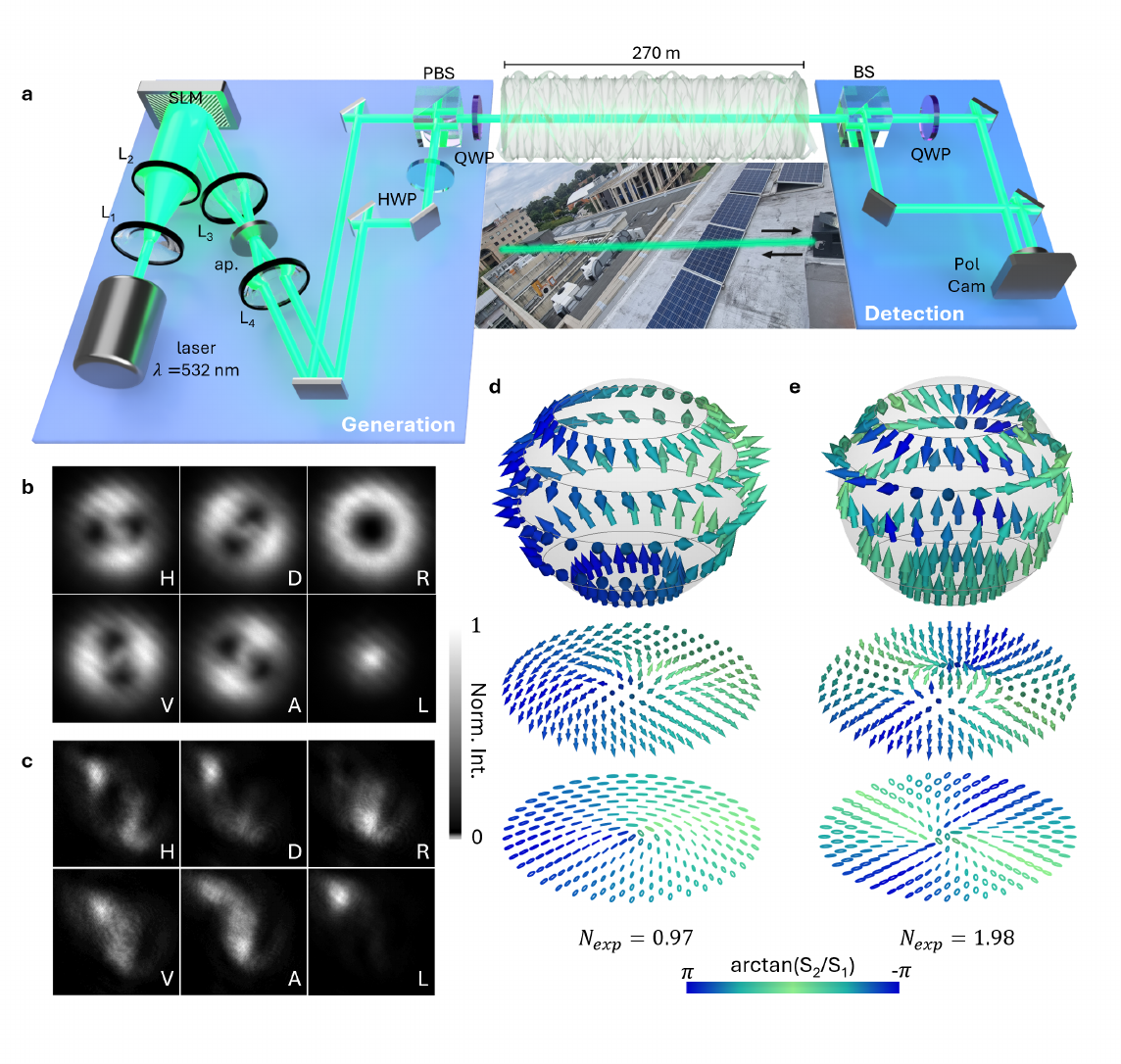}
    \caption{\textbf{Generation and detection through a free space link.} \textbf{a}. Optical skyrmions were generated using complex amplitude holograms to generate two scaler modes. These modes were vectorially combined with a Mach-Zehnder interferometer and transmitted through a 270~m free space optical link. At the receiver, Stoke polarimetry was performed on the aberrated beam using a 50:50 beam splitter, quarter-wave plate and a polarisation sensitive camera \textbf{b}. Polarisation intensity projections for all 6 polarisation states captures simultaneously for a skyrmion with $N=1$ before the beam had propagated through the free space link. \textbf{c}. Polarisation intensity projections for all 6 polarisation states captures simultaneously for a skyrmion with $N=1$ after the beam had propagated through the free space link. \textbf{d}. State of polarisation, Stokes vector texture and mapping to the spatial sphere for a unaberrated skyrmion beam with $N=1$ captured using the detection scheme \textbf{e}. State of polarisation, Stokes vector texture and mapping to the spatial sphere for a unaberrated skyrmion beam with $N=2$ captured using the detection scheme}
    \label{fig:ExpSetup}
\end{figure*}

Optical skyrmions were generated experimentally with the setup shown Figure \ref{fig:ExpSetup} \textbf{a}. Here a laser of wavelength $532$~nm was shaped into various vectorial beams in the form of Equation \ref{eq:RL vector beam} with $l_1 = 1,\,2$ and $3$ and $l_2=0$ using a spatial light modulator (SLM) and a modified Mach–Zehnder interferometer, to generate optical skyrmions with $N=1,\,2$ and $3$, respectively. These beams were then transmitted across a 270~m long outdoor free space link. At the receiver, a polarisation sensitive camera and 50:50 beam splitter are used to measure the horizontal (H), vertical (V), diagonal (D), antidiagonal (A), right-circular (R) and left-circular (L) polarisation intensity projections in a single shot similar to methods implemented in Refs. \cite{cox2023real,peters2023invariance}. Example projections taken for a skyrmion with $N=1$ before and after propagation through the free space link are shown in \ref{fig:ExpSetup} \textbf{b} and \textbf{c}, respectively. The simultaneous measurement of these projections is essential to ensure that the full state of polarisation (SoP) of the beam is determined for a given atmospheric configuration within the coherence time of the channel. 
Figure \ref{fig:ExpSetup} \textbf{d} and \textbf{e} shows the experimentally reconstructed state of polarization (SoP), Stokes vector texture and the mapping onto the spatial sphere for optical skyrmions with $N=1$ and $N=2$ before propagating through the free space link. Here we clearly see how each distinct topology wraps a different number of times around the Poincar\'{e} sphere and we can qualitative evaluate the topology. For $N=1$, the polarisation states move from horizontal (blue) to vertical (green) only once and then back to  horizontal as one moves azimuthally around the optical axis. This occurs twice in the SoP for $N=2$. Quantitatively, the experimentally computed sphere coverage of $N_{ \rm exp} = 0.97$ and $N_{ \rm exp}=1.98$ calculated using Equation \ref{eq:skyrmion wrapping} is also extremely close to the encoded values indicating the successful generation and detection of the desired topologies before being exposed to turbulence.\\

\begin{figure*}[t!]
    \centering
     \includegraphics[width=0.9\linewidth]{ 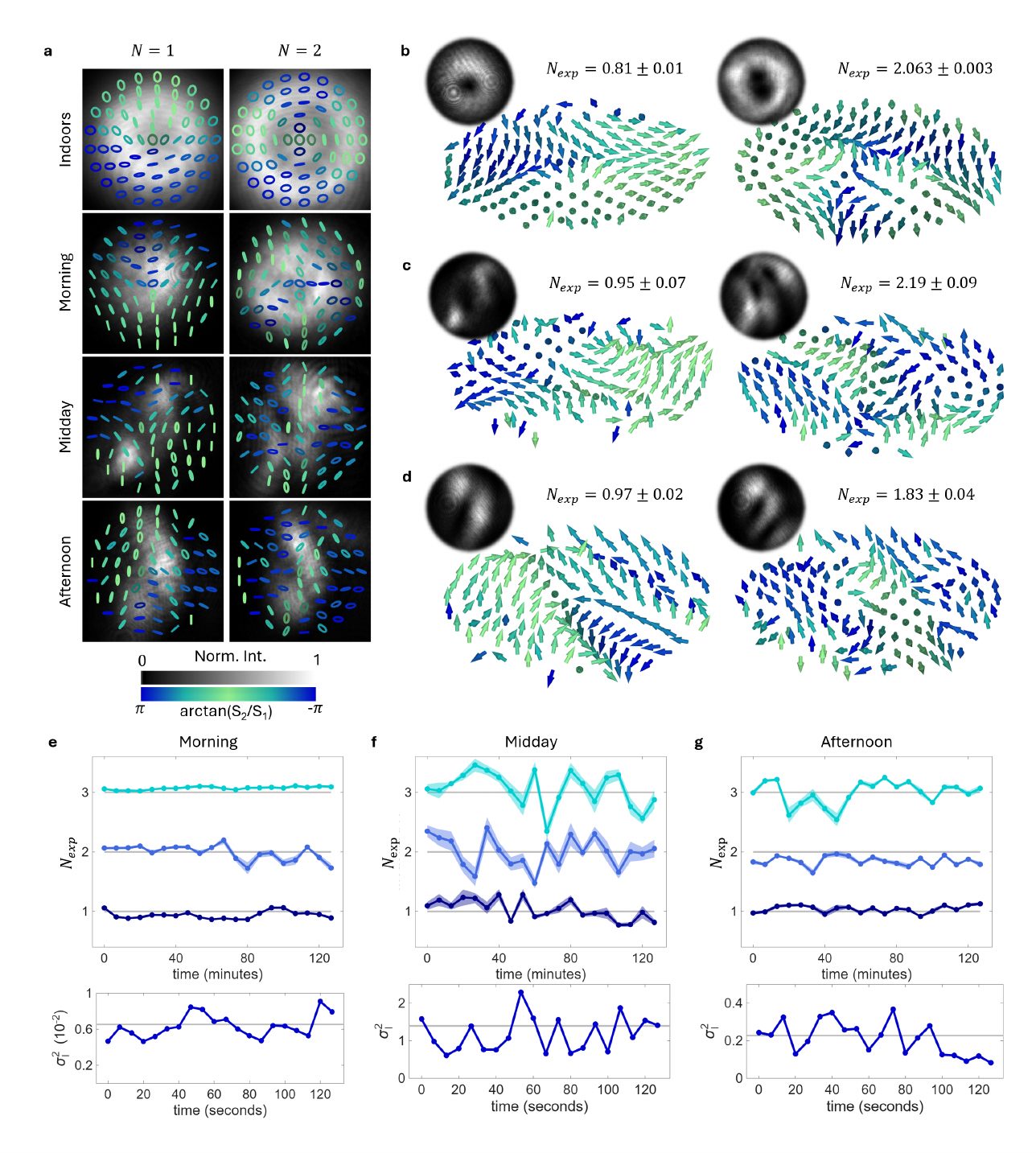}
    \caption{\textbf{Robustness through free space link.} \textbf{a}. Total intensities and the measured state of polarisation for skyrmion with $N=1$ and $N=2$ after propagating indoors and through the free space link in the morning, midday and late afternoon. Experimentally measured Stokes vector textures (main) and corresponding OAM scalar component (inset) for skyrmions transmitted in the \textbf{b}. morning, \textbf{c}. at midday and \textbf{d}. in the late afternoon. The measured skyrmion number for encoded numbers of $N=1$, 2 and 3 and the measured scintillation index $\sigma_I^2$ measured over 120~s in the \textbf{e}. morning, \textbf{f}. at midday and \textbf{g}. in the late afternoon. }
    \label{fig:AvgResults}
\end{figure*} 

\noindent \textbf{Robustness through real-world free space link.} Figure \ref{fig:AvgResults} \textbf{a} shows experimentally obtained intensities and SoPs for the skyrmions with $N=1$ and $N=2$ indoors (no turbulence), and through the free space link in the morning, midday and late afternoon. Morning weather conditions result in the least severe distortions, as seen by the only slight changes in the intensity profiles as compared to the indoor case, with the SoP roughly maintaining its cylindrical symmetry through the channel. These slight variations can be attributed to the phase distortions induced by the channel. Over and above this broken symmetry, we also see that most of the polarisation states have changed in propagation. This is a result of a difference in the Gouy phase of the component scalar beams \cite{baumann2009propagation}. In addition to a radius of curvature term, each LG beam gains a phase of $e^{i\psi(z)}$, where $\psi(z) = (2p + |l| +1)\arctan(z/z_R)$ and where $z_R$ is the Rayleigh length of the beam. This phase depends on the mode order and thus the $l$-index of the scalar mode. Since the component modes of the vector beam must have differing $l$-indices to form the skyrmion mapping, the scalar components of each beam will accumulate phase at different rates, effectively having different propagation constants through free space. As a result, the polarisation structure will change in propagation, even in the absence of a complex channel such as atmospheric turbulence. This effect does not affect the skyrmion mapping and only changed the polarisation texture. The data taken in the late afternoon shows much more severe distortion in both amplitude and polarization. Here, we see that the intensity has lost its cylindrical symmetry, with highly distorted profiles. The SoP is also highly aberrated, with rapid variations in polarisation across the transverse plane as compared to the smooth transitions seen in the SoP of the indoor data. The midday data shows the most severe distortion, with separate lobes developing in the intensity profiles of the beams and far more rapid variations in the SoPs. The change in the severity of the distortion is easily explained by the changes in weather during the day. Morning temperatures start off quite low, with cool air and surfaces resulting in small temperature gradients and minimal turbulence. Midday sees the highest temperatures, heating up surrounding buildings, pavements and roads and creating large temperature gradients ideal for creating very strong turbulence. By late afternoon these surfaces begin to cool, leading to a decrease in the turbulence strength. 

With a qualitative understanding of the  distortion experienced at each of the three times of day, we may now quantify the topology after the beams have passed through the link at these different times. Figure \ref{fig:AvgResults} \textbf{b} shows the computed Stokes vector texture for the morning data shown in Figure \ref{fig:AvgResults} \textbf{a}. The insets show the intensity of the scalar mode carrying OAM. We see that while the vector textures differ noticeably from the textures plotted in Figure \ref{fig:ExpSetup} \textbf{c} and \textbf{d}, the topological features are maintained. The Stokes vector at the centre still points upwards, and the vectors rotate downwards as one move radially outwards. For the $N=1$ skyrmion, we see that the vectors rotate once around as we move azimuthally, indicating one wrapping around the Poincar\'{e} sphere and illustrated with the colourmap as moving from blue to green and back to blue only once. For the $N=2$ skyrmion, this rotation occurs twice indicating two wrappings around the Poincar\'{e} sphere. These features agree well with the computed wrapping numbers of $N_{ \rm exp}=0.81\pm0.01$ and $2.063\pm0.003$ respectively. Figure \ref{fig:AvgResults} \textbf{c} shows the computed Stokes vector texture for the midday data shown in Figure \ref{fig:AvgResults} \textbf{a}. Due to the severe distortion experienced at this time of day, the trends discussed previously are not as easy to visually discerned from the plotted textures. However, we still see the single wrapping and double wrapping for $N=1$ and $N=2$, respectively as one moves azimuthally. We also see good agreement between the encoded and measured skyrmion numbers of $N_{ \rm exp}=0.95\pm0.07$ and $2.19\pm0.09$, indicating robustness of the topology through even the most extreme cases tested.  Figure \textbf{d} shows the computed Stokes vector texture for the afternoon data shown in Figure \ref{fig:AvgResults} \textbf{a}. Here the distortion was milder than at midday but stronger than in the morning, with noticeable distortion in the vector textures. Despite this, we can still visually discern the single azimuthal wrapping for skyrmion $N=1$ and double wrapping for $N=2$. The experimentally evaluated values of the distorted beams of $N_{ \rm exp}=0.97\pm0.02$ and $N_{\rm exp}=1.83\pm0.04$, again very closely match the encoded values. 

We have established that the skyrmion number remains unchanged under specific instances of atmospheric turbulence but it is important to study the robustness of this topology as the medium changes. We show the measured skyrmion number over approximately 120 second in Figure \ref{fig:AvgResults} \textbf{e}, \textbf{f} and \textbf{g} in the morning, at midday and late afternoon respectively. We also show the measured scintillation indices $\sigma_I^2$ for the respective times of day, calculated using the variation of the on axis intensity $I(t)$ of left-circularly polarised component ($l_2=0$) of the transmitted beam according to \cite{peters2025structured}
\begin{equation}
    \sigma_I^2 = \frac{\langle I(0,0,t) \rangle^2}{\langle I^2(0,0,t) \rangle}-1\,,
\end{equation}
where $\langle\cdot\rangle$ represents the average over time. Each data point shows the average measured skyrmion number over five sequential 2~ms measurements to allow for a measurement uncertainty to be determined. There were 6 second intervals between each data point and the shaded region indicating the measurement uncertainty. Atmospheric conditions in the morning showed an average scintillation index of $\sigma_I^2 \approx 0.007$, comparable to strengths seen in other free-space link experiments \cite{mcdonald2025field}. This supports the qualitative evaluation made earlier, with only mild aberration seen in the beams' intensity profiles and SoPs. The skyrmion number for each of the tested topologies, $N=1,\, 2$ and $3$, remains fairly consistent over the measurement period, with only small fluctuations in the $N=2$ topology in  the second half of the measurement. The midday measurements show an average scintillation index of $\sigma_I^2\approx1.4$. This can be considered a very high value and corresponds well with the results in \ref{fig:AvgResults} \textbf{a}. The measured skyrmion number over the measurement period is not as steady as seen in the morning data, and fluctuates noticeably around the encoded value. However, despite these fluctuations, the value of each topology at almost every point in time is discernable from every other point and close enough to the encoded value to determine what the initially encoded topology was. It is worth noting that when the wrapping number is reported, it is calculated using Equation \ref{eq:skyrmion wrapping} which gives the coverage of the Poincar\'{e} sphere. This coverage calculation can be prone to errors, especially in the presence of background noise, limited ROI and nonsymmetrical beam profiles. All of these factors are present in our measurements and impossible to completely mitigate. Despite these challenges, we are able to easily distinguish between various encoded topologies even in the most extreme cases through the real-world free space link. The afternoon data shows smaller variation over the measurement period for all three encoded topologies as compared to the midday data but more severe than the morning data, with an average distortion strength of $\sigma_I^2 \approx 0.23$. This again aligns with the qualitative observations discussed earlier, over this measurement period. Here we again see that the measured skyrmion number fluctuates over the course of the measurement, but still remains close to the encoded value, allowing us to easily identify what the transmitted topology was through the distorting channel.  \\[0.3cm]

\begin{figure*}[htpb!]
  \centering
   \includegraphics[width = 0.9\linewidth]{ 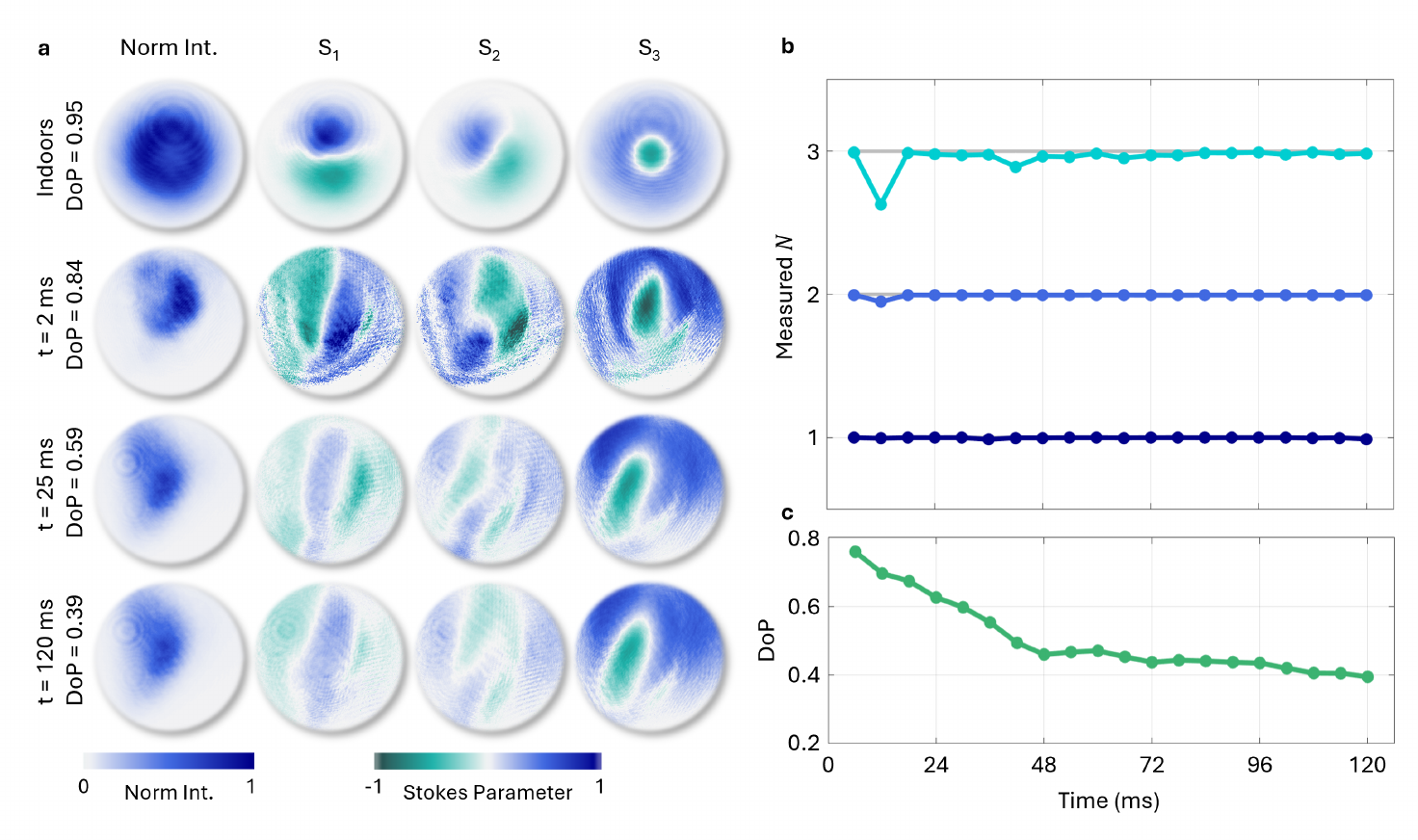}
    \caption{\textbf{Resilience to depolarisation and decoherence.} \textbf{a}. The measured intensity and locally normalised Stokes parameters of a skyrmion beam with $N=1$ measured indoors (top row), after averaging for 2~ms (2nd row), after averaging for 25~ms (3rd row) and after averaging for 120~ms (fourth row). \textbf{b}. Line plot showing the measured skyrmion number for encoded numbers of $N=1$, 2 and 3  against the time over which the measurement was averaged. \textbf{c}. Line plot showing the measured degree of polarisation against the time over which the measurement was averaged. }
   \label{fig:DePolData}
\end{figure*}

\noindent \textbf{Resilience to depolarisation and decoherence in turbulence.} So far, we have discussed the robustness of optical skyrmions to atmospheric phase distortions and the scintillations that results from these. The measurements were taken over short measurement times, and did not allow the channel to significantly evolve during the acquisitions time for each data point. In the context of classical optical communications, this robustness is vital for reliable and effective data transmission. However, skyrmions have also recently been realised in the quantum domain, for both single photon states \cite{ma2025nanophotonic} and in hybrid entangled biphoton states \cite{ornelas2024non}, promising increased security and data protection for quantum communication. Measurement of quantum states requires significantly longer measurement times, with even the most state of the art techniques requiring measurement times on the order of $\approx10^2$ seconds \cite{ndagano2020imaging}. Typical atmospheric coherence times are on the order of 10-100~ms\cite{fried1990greenwood}. Therefore, quantum states sent through turbulence will undergo multiple realisations of the channel over the course of a single measurement, leading to additional deleterious effects caused by depolarisation and decoherence. Several experiments have simulated the effects of these channels on optical skyrmions, with promising results demonstrating possible resilience \cite{de2025quantum}. Here, we have a unique opportunity to study this phenomenon in a real-world channel. Using classical light, we can easily probe the effects of depolarisation and decoherence, with the results being directly transferable to quantum skyrmions due to the close mathematical analogy between hybrid entangled biphoton states and classical vector beams \cite{ndagano2017characterizing,shen2022nonseparable,peters2023spatially}. We show our results in Figure \ref{fig:DePolData}. The first row of Figure \ref{fig:DePolData} \textbf{a} shows the normalized total intensity and Stokes parameters for a skyrmion with $N=1$ under indoor conditions, with no turbulence. We see that the Stokes parameters are unaberrated, and the degree of polarisation (DoP) is close to 1, meaning the beam is almost fully polarised and coherent. The second row of Figure \ref{fig:DePolData} \textbf{a} shows the Stokes parameters for the same beam, but now after a 2~ms exposure time through the turbulence channel. The Stokes parameters are distorted, but the beams DoP is still relatively high, having decreased primarily because of the presence of ambient, thermal light such as sunlight. The third row of Figure \ref{fig:DePolData} \textbf{a} shows the Stokes parameters for a beam after averaging for 25~ms. Here we can see that the magnitudes of the parameters have decreased as compared to the second row. The DoP of this state is 0.59 , indicating that a portion of the field measured is now partially incoherent/depolarised. However, the computed skyrmion number still remains close to the encoded value of $N=1$. The fourth row of Figure \ref{fig:DePolData} \textbf{a} shows the Stokes parameters for the same beam after averaging for even longer, 120~ms. Here we see that the amplitudes of the Stokes parameters have decreased even further, indicating even more severe depolarisation. This is further supported by the calculated DOP for this state, which is 0.39. The measured wrapping number for this beam is $N=0.98$, still close to the encoded value. Figure \ref{fig:DePolData} \textbf{b} shows the measured skyrmion number for $N=1,2$ and $3$ as the measurement was averaged over longer and longer time periods. Figure \ref{fig:DePolData} \textbf{c} plots the degree of polarisation (DoP) over averaging time, indicating the severity of the effect of the depolarisation during the course of the measurement. We see that the skyrmion number remains almost perfectly unchanged for all three encoded values from the initial exposure time of 2~ms all the way up to the a total of 240~ms (2~ms exposure per image averaged over 120 images). This is despite the fact that the DoP drops from $\approx0.84$ to $\approx0.39$, a significant decrease over the course of the measurement. This shows that optical skyrmions are immune to the effects decoherence and depolarisation caused by real-world atmospheric turbulence.

\begin{figure*}[htpb!]
  \centering
   \includegraphics[width = \linewidth]{ 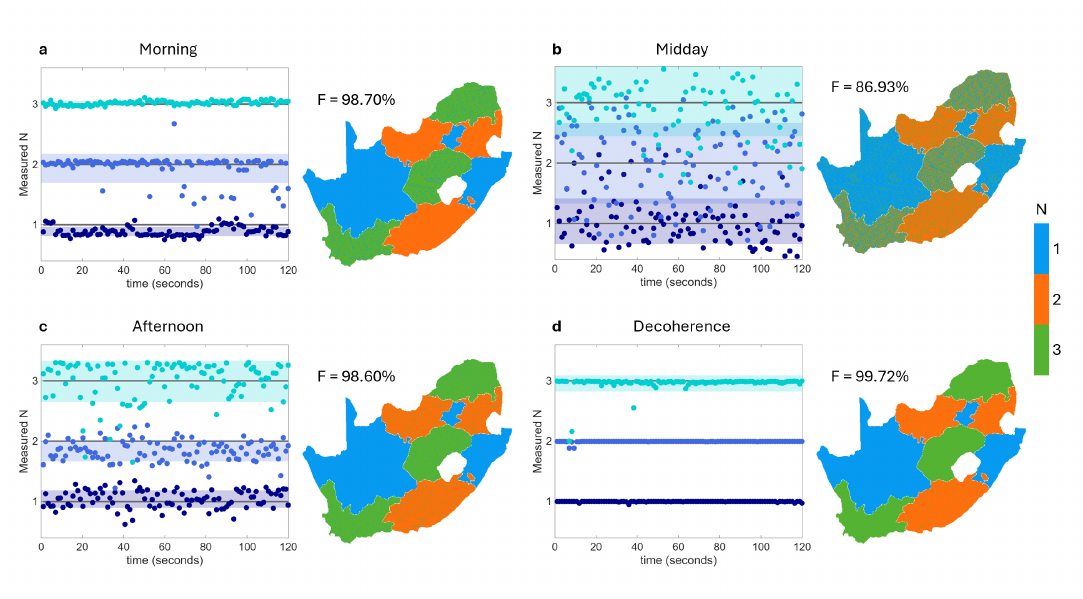}
    \caption{\textbf{Robust information encoding.} The measured skyrmion numbers over the measurement period and the reconstructed image of the transmitted image performed \textbf{a} in the morning under mild distortion, \textbf{b} at midday during the most severe distortion conditions, \textbf{c} in the late afternoon under moderate distortion and \textbf{d} under the effects of increasing depolarisation. }
   \label{fig:Comms}
\end{figure*}

\noindent \textbf{Robust information transfer in atmospheric turbulence.} We have now seen that optical skyrmions are resilient to a wide variety of turbulence channels. One promising applications for this is robust information encoding through free space. Atmospheric turbulence typically distorts more common forms of structured lights, leading to severe crosstalk if no compensation techniques are used to mitigate its effects. In contrast, the wrapping number of the generated optical skyrmions typically remains unchanged through these channels, meaning it can safely encode data through the link. To demonstrate this experimentally by way of visual example shown in Figure \ref{fig:Comms}. Here, we encode a 3-colour, $512\times560$ image of South Africa using topology, where each $N$ corresponds to a particular colour, with $N=1,2$ and $3$ corresponding to blue, orange and green respectively. We then transmit the image one pixel at a time over a time interval of 2~ms with the appropriate topological state being transmitted through the channel and measured at the receiver. The received image is constructed by measuring the skyrmion number of the received state. If fractional values are measured, they are rounded off to the nearest integer. Figure \ref{fig:Comms} \textbf{a} shows the results taken in the morning, with the line graph showing examples of the individual measured skyrmion numbers during measurement period and the shaded region indicating the mean plus/minus one standard deviation. As before the variation in the measured number during the morning is minor, with only a few data points for $N=2$ deviating noticeably. The reconstructed image maintains its structure, with each province in the map maintaining a uniform, unaltered colour indicating that the information was transmitted with an almost perfect fidelity of $98.70\%$. Figure \ref{fig:Comms} \textbf{b} shows results taken at midday, exhibiting significant deviations from the encoded value due to the significant turbulence strength experienced during this time of day. We see this quantitatively with the reconstructed image having a fidelity of $86.93\%$. We do see some noticeable discolouration in the reconstructed image as a result of this, especially for the higher order topologies of $N=2$ and $3$. However, the various regions in the reconstructed image are still discernable from each other and the initial encoded colour is still easily identifiable. Figure \ref{fig:Comms} \textbf{c} shows the results taken in the afternoon. We see that the variation is higher than in the morning but less then midday. There are some slight errors for the $N=2$ topology, evident by the presence of some green pixels in regions which are orange in the encoded image, but overall the image integrity was well maintained through the channel, seen with a fidelity of $98.60\%$. Figure \ref{fig:Comms} \textbf{d} shows the same data but under the condition where the measurement is averaged over 240 ms  in turbulence to allow for there to be significant depolarisation in the same regime as Figure \ref{fig:DePolData}. In this case the fidelity is extremely high at $99.72\%$. We see almost no variation in the measured $N$ due to fact that averaging the measurement also averages over the effects of the ambient light and shot noise of the camera, reducing its deleterious effects on the   wrapping number calculation. Overall, we see that topology make a useful alphabet for transmitting information through the free space link, only showing a noticeable drop in performance in the most extreme cases, where the distortion is severe. However, even in this regime, without any measurement or probing of the channel or pre- or post compensation, the image is able to be transmitted with a reasonably high fidelity.\\

\section*{Discussion}
Experiments so far have primarily focused on the generation of optical skyrmions and other form of topological light, with only a few recent works investigating their robustness through complex channels. These investigations, while valuable, were performed in controlled laboratory settings.  In this work, we have presented the first experimental evidence of the robustness of optical skyrmions through a real-world turbulent channel. We have shown that the topology is preserved under the phase, amplitude and polarization distortions imparted onto any light field that passes through such a channel. This was demonstrated over a wide variety of conditions, ranging from morning to midday and late afternoon, where the strength of the experienced turbulence varied from $\sigma_I^2 \approx 0.007$ to $\sigma_I^2 \approx 1.4$. We also demonstrated the invariance of the skyrmionic topology to the decoherence and depolarization effects that are typically observed when averaging over the time varying effects of the turbulence channel. Here, the skyrmion number remains unchanged even when the degree of polarization dropped to as low as 0.39. In both cases, the robustness was achieved by harnessing properties of the light itself in order to generate structures that are invariant to actions of the channel. This is contrast to many other proposed solutions to the issue of atmospheric turbulence, which either require a form of pre- or post-compensation such as adaptive optics or require a probe or measurement of the channel, such as finding its invariant spatial. Here, the invariant passes through the channel as if it was transparent, and no knowledge of the channel is required. The use of topology also has benefits over similar approaches to finding and leveraging invariants, such as using the non-separability of vectorial beams, which require the channel to be unitary. While this is the case for the phase and amplitude distortions, this does not occur when one experiences decoherence/depolarisation \cite{roux2011infinitesimal}. As we have shown in this work, the topology is robust not only through a unitary channel, but also when more exotic transformations are considered. 

It is worth noting that in this investigation, we made use of a polarisation camera and Stokes polarimetry \cite{singh2020digital} to measure the skyrmion number. In the absence of a dedicated topological detector, such means are currently the best available tools for this purpose. However, such approaches are also prone to errors and fluctuations when computing the wrapping number, especially in the presence of high ambient noise such as sunlight and differing path lengths for different polarisation projections in experiment. While these factors can be mitigated to a certain degree, they can never be full removed resulting in noticeable fluctuations in the measured data. Interest in vectorial light in general and optical skyrmions in particular has driven research into more accurate measurement schemes \cite{mcwilliam2023topological} and dedicated polarisation projective metasurfaces \cite{li2025disorder} for faster and more accurate measurement techniques. We note that despite its simplicity and vulnerability to noise, the measurement setup used here is still able to obtain accurate and reliable measurements of the wrapping number through the free space link even when the distortion strength is quite severe. We hope this work then demonstrates the promise of topological light for use in such media and motivates further research for even better tools to accurately retrieve the topological characteristics of structured light in these highly topical applications. 

In this work, we generated optical skyrmions using specific combinations of LG beams. This firstly means that higher order topologies experienced stronger turbulence strengths, due to the inherent scaling of LG transverse beam size with OAM \cite{cocotos2025orbital}. As such, a fair comparison of robustness of topologies with different $N$ cannot be made with these results and is an open questions that merits further investigations. Additionally, skyrmions in this work were made by setting $l_2=0$ and modulating $l_1$ to change $N$, Different combinations of $l_1$ and $l_2$ can result in the generation of beams with different underlying spatial modes, but the same $N$. Optimizing the underlying geometry of topological light to increase its robustness has already sown promise for optical knots \cite{pires2025stability}. It may therefore be of interest for future studies to investigate whether other vectorial combinations of LG beams, forming skyrmions with the same $N$ tested here, could offer even greater resilience to the aberrations induced by complex channels.

Robustness to phase and amplitude distortions resulting from turbulences, as well as the resilience to decoherence, is of immense interest to the field of free space optical communication, where the search for a reliable, resilient and high dimensional encoding alphabet has the potential to dramatically increase bandwidths and data transfer speeds. We have demonstrated in this work not only the robustness of these topological states to this highly complex medium, but also their ability to faithfully transmit data through such channels with ease. We envision that our results will pave the way for a new approach to see and transport information through distortion by harnessing light's innate, invariant properties for fast, reliable and efficient communication. 

\section*{Methods}

\noindent \textbf{Experimental Setup.} A diagram of the experimental setup used in this work can be seen in Figure \ref{fig:ExpSetup} \textbf{a}. Here, a green laser beam ($\lambda = 532$~nm) was expanded using a 10$\times$ objective lens $L_1$ and collimated with a second lens $L_2$ of focal length $f_2 = 500$~mm. This beam was then incident on a PLUTO-VIS HoloEye spatial light modulator (SLM). The SLM screen was split in half, with each half generating one component scalar mode of the final vectorial beam using a complex amplitude modulation scheme \cite{arrizon2007pixelated},
\begin{equation}
    H(x,y) = J^{-1}_1 (A(x,y)) \sin[\phi(x,y) + 2\pi(G_x x + G_y y) ] \,,
\end{equation}
where $A(x,y)$ and $\phi(x,y)$ is the amplitude and phase of the desired field respectively, $J^{-1}_1$ is the inverse Bessel function of the first kind and $G_{x(y)}$ are the grating frequencies in the horizontal and vertical directions. Both beams are created with an embedded Gaussian beam waist of $w_0 = 0.665$~mm and are sent through a 4f imaging system consisting of lenses $L_3$ and $L_4$ (focal lengths $f_3=f_4=300$~mm) with an iris in between acting as a spatial filter to isolate the first diffraction order of the hologram where our encoded scalar fields are located. These beam are propagating, but not coaxial, and are thus sent through a modified Mach-Zehnder interferometer, with a half-wave plate (HWP) in one arm, to form the desired vectorial beam with the aid of a polarizing beam splitter (PBS). Before beam transported through the free space link, the vectorial beam is sent through a quarter-wave plate (QWP) to convert the scalar component modes to the right and left circular polarization basis. It is then sent through two further telescope systems, each providing a further 3$\times$ magnification such that the beam transmitted through the free space link has an embedded Gaussian beam waist of $w_0 = 6$~mm. The telescope systems are constructed such that the beam focusses on a flat mirror located on the top of the opposite building, located approximately 135~m from the last telescope system, for a round trip distance 270~m. Upon returning, the beam is demagnified by a telescope system with a $0.125\times$ magnification and then incident on a 50:50 beam splitter. One port of the beam splitter leads directly to one half of the sensor of a polarisation sensitive camera (Allied vision Mako G-508B POL) to measure the four linear polarization states: horizontal, vertical diagonal and antidiagonal. The other port of the beam splitter passes through a quarter-wave plate (QWP) before being incident on the other half of the camera sensor to measure the right and left-circular polarisation states.\\

\noindent \textbf{Experimental determination of the skyrmion number} The wrapping number of an optical skyrmion is typically calculated according to Equation \ref{eq:skyrmion wrapping}. In this work, we use an alternative formulation proposed by McWilliam \textit{et al.} \cite{mcwilliam2023topological}, which makes use of a contour integral over complex polarisation fields. Primarily, we make use of their result which expresses the skyrmion number as follows,
\begin{equation} \label{eq:line_int}
     N = \frac{1}{2} \left( \sum_{j} S_{z}^{(j)} N_{j} - \bar{S_{z}^{\infty}} N_{\infty} \right) \,,
\end{equation}
where $N_j$ is the charge of individual phase singularity at position $j$ in the field $S_x + iS_y $,  $S_z^{(j)}$ is the value of the Stokes parameter $S_z$ at the point $j$, $N_{\infty}$ is the result of the contour integral at infinity and $S_z^{(\infty)}$ is the value of the Stokes parameter $S_z$ as $r\rightarrow \infty$. 

The Stokes parameters of the generated optical skyrmions were experimentally measured using six polarisation intensity projections,
\begin{eqnarray} \label{eq:stokesS0}
    s_{0} &=& I_{H} + I_{V} \\
\label{eq:stokesS1} 
    s_{1} &=& I_{H} - I_{V}\\
\label{eq:stokesS2}
    s_{2} &=& I_{D} - I_{A}\\
\label{eq:stokesS3}
    s_{3} &=& I_{R} - I_{L} \,.
\end{eqnarray}
where the subscripts H, V, D, A, R and L represent horizontal, vertical, diagonal, antidiagonal, right circular and left circular polarisations respectively. Equation \ref{eq:line_int} requires the locally normalised Stokes parameters $S_j$ which were computed from the experimentally obtained Stokes parameters $s_j$ according to,
\begin{equation}
    S_j =\frac{s_j}{\sqrt{s_1^2+s_2^2+s_3^2}} \,.
\end{equation}
In order to determine the location and charge of the singularities $N_j$, the polarisation fields $P$ were calculated using,
\begin{equation} \label{eq:polphase1}
P = S_{x} + iS_{y}\\
\end{equation}

\textbf{Post-processing and data analysis.} Due to the presence of ambient noise and the highly complex structures of the beam being measured, it was necessary to perform several post-processing steps to accurately recover the skyrmion number. The steps taken were informed by the physics of the system, being chosen as to only smoothly deform the topology of the received beam, while allowing us to average out or exclude noisy contributions. The first step is averaging over multiple rotations of the Poincar\'{e} sphere. Such rotation may change the geometry of the measured state, but must leave the topological invariant unchanged. This is possible because Equations \ref{eq:line_int} and \ref{eq:polphase1} are able to accept any three Stokes parameters, so long as they are ordered correctly and orthogonal on the Poincar\'{e} sphere. With the use of a simple rotation matrix, one may rotate the measured Stokes parameters an infinite number of times, each time forming a slightly different measured field. Averaging over these rotations helps eliminate the random fluctuations induced by noise. In practice the measured number does change as one rotates the Stokes parameters, but by taking the average over 300 individual rotations, we are able to eliminate a significant portion of the noise effects and retrieve a more faithful value for the wrapping number.

FOr each rotation of the Stokes parameters, the locations and charges of the individual singularities  in each polarisation field $P$ were computed using a numerical equivalent to the curl $\nabla \times$ operation proposed in Ref \cite{chen2007detection} termed the circulation $D$. The circulation is defined as,
\begin{eqnarray}
    D^{m,n} = &\frac{d}{2}& ( G_x^{m,n} + G_x^{m,n+1} +G_y^{m,n+1} + G_y^{m+1,n+1} \nonumber \\ &-  &G_x^{m+1,n+1} - G_x^{m+1,n} -G_y^{m+1,n} \nonumber \\
    &-& G_y^{m,n})\,.
\end{eqnarray} 
Here, $D^{m,n}$ represents the value of the circulation of the pixel in the $n$-th row and $m$-th column. $G_x^{m,n}$ and $G_y^{m,n}$ are the phase gradient in the horizontal and vertical direction of the pixel in the $n$-th row and $m$-th column, respectively and $d$ is the pixel size. Typically, the circulation will return a 0 value if there is no singularity at that pixel and a nonzero value if there is. The magnitude of the circulation indicates the charge of the singularity and the sign indicates the direction/handedness of the singularity. These values were then substituted into Equation \ref{eq:line_int} to calculate the skyrmion number.

We also implemented a low-pass 2D Gaussian filter with smoothing kernel. The standard deviation of this kernel was $\sigma=1498$~m$^{-1},1981$~m$^{-1}$ and $1653$~m$^{-1}$ for the morning, midday and afternoon data respectively. This filter is not universally a smooth deformation, but rmeains smooth up to reasonable numerical limits \cite{wang2024topological}. A final step was the use of an intensity base threshold for the data, which changed over the course of the day to different levels of background noise from the environment. The thresholds were 0.02, 0.035 and 0.025 of the maximum intensity for the morning, midday and afternoon data respectively. Such a threshold is necessary, because light measured from ambient sources such as the sun and shot noise in the camera is unpolarised and so cannot be faithfully assigned a polarisation state. No Stokes vector can be assigned to these pixels in low intensity regions, ans so they must be excluded from the calculation of the topological invariant. \\ 

\section*{Acknowledgments}
C.P. would like to thank Pedro Ornelas for useful discussions regarding this project and assisting with code used in the data analysis.
\paragraph*{Funding:}
A.F. would like to acknowledge support from the South African National Research Foundation/CSIR Rental Pool Programme and the South African Quantum Technology Initiative.  E.B. would like to acknowledge the support of IdEx University of Bordeaux and its Grand Research Program GPR LIGHT
\paragraph*{Author contributions:}
C.P. and A.D. designed and built the experiment. V.H. and C.P. performed the experiment and curated the data. C.P. analysed the data and wrote the first draft of the manuscript. All authors discussed the results and contributed the manuscript. A.F. conceived of the idea. E.B, M.C. and A.F. supervised the project. 
\paragraph*{Competing interests:}
There are no competing interests to declare.
\paragraph*{Data and materials availability:}
The data underlying the results reported in this work can be found online at   \href{https://github.com/CadeRPhysics/SkyrmionsFreeSpaceLink}{https://github.com/CadeRPhysics/SkyrmionsFreeSpaceLink}.

\bibliography{report} 
\bibliographystyle{sciencemag}

\end{document}